# Raman-scattering study of the phonon dispersion in twisted bi-layer graphene

*J. Campos-Delgado[1\*], L. G. Cançado[2], C.A. Achete[3], A. Jorio[2], J.-P. Raskin[1]*

[1] ICTEAM, Université catholique de Louvain, Louvain-la-Neuve, 1348, Belgium

[2] Departamento de Física, Universidade Federal de Minas Gerais, Belo Horizonte, MG, 30123-970, Brazil

[3] Divisão de Metrologia de Materiais, INMETRO, Xerém, RJ, 25250-020, Brazil

**Corresponding Author**

\* Author to whom correspondence should be addressed: jessica.campos@uclouvain.be

**Abstract**

Bi-layer graphene with a twist angle $\theta$ between the layers generates a superlattice structure known as Moiré pattern. This superlattice provides a $\theta$-dependent $\boldsymbol{q}$ wavevector that activates phonons in the interior of the Brillouin zone. Here we show that this superlattice-induced Raman scattering can be used to probe the phonon dispersion in twisted bi-layer graphene (tBLG). The effect reported here is different from the broadly studied double-resonance in graphene-related materials in many aspects, and despite the absence of stacking order in tBLG, layer breathing vibrations (namely ZO' phonons) are observed.

**Introduction**

The double-resonance (DR) Raman scattering mechanism has been largely explored in graphene-based materials [1-3]. Having being studied in graphite [4-8], graphite whiskers [9], carbon nanotubes [10-13], graphene [14,15], nanographite and amorphous carbons [16,17], the DR Raman is a striking effect that allows one to probe the interior of the phonon dispersion relation using visible light by changing the excitation laser energy. It was introduced by Thomsen and Reich[4] to explain the observation of the defect-induced D band in graphitic materials, the one phonon Raman scattering process being always constrained to the existence of defects to break the $q \cong 0$ momentum selection rule. Furthermore, the DR mechanism selects the modulus of the phonon wavevectors $|\boldsymbol{q}|$, and the observed peaks give an average of the phonon frequencies observed around the high symmetry $\Gamma$ and $K$ points in the graphene Brillouin zone [5], pounded by the electron-phonon interaction which is wavevector direction dependent [15].

Bi-layer graphene with a mismatch angle θ between the layers, named twisted bi-layer graphene (tBLG), generates a modulation which activates one phonon Raman processes without requiring defects [18-20]. This effect has been observed recently by different groups [18-23], and here we provide a unifying picture that accounts for all Raman peaks observed in tBLG, showing that the superlattice-induced Raman process allows us to probe the phonon dispersion of tBLG. Different from previously studied DR in graphene-based systems, in tBLG the phonon dispersion can be probed without changing the excitation laser energy, without the presence of defects or higher-order processes, and the process is $q$-direction selective.

**Results and Discussion**

Our sample is composed of a mono-layer film where bi- and multi-layer islands can be observed. These islands sitting on mono-layer graphene have in general a multi-lobe star morphology; the lobes constituting the second layer and, in the center, 3 or more stacked layers can be identified by optical contrast when using an optical microscope. Recently, it has been proven that the electronic properties of twisted bi-layer graphene combined with its reflection on a 100 nm-thick $SiO_2$/Si substrate make twisted bi-layer graphene (9°<θ<16°) visible through the appearance of blue, yellow or pink/red colorations [24,25]. We have transferred our samples to Si/$SiO_2$ (100 nm) substrates and identified bi-layer lobe areas with different colorations. We have recorded a large number of Raman measurements on different pink-, blue- and yellow- colored lobe areas, representative zones of measurement are shown in Figure 1 marked by pink, blue and yellow circles, respectively. This allowed us to probe twisted bi-layer samples with rotation angles from 9° to 16° using Raman spectroscopy.

A detailed inspection of our Raman data from these bi-layer zones confirms variations of rotational angles, since different frequencies between 1450 cm$^{-1}$ and 1525 cm$^{-1}$ were found for the R bands, as shown in References [18-20]. However, we observed additional peaks at frequencies between 100 and 900 cm$^{-1}$. Figure 2 shows representative spectra of the samples, where the peaks originated from tBLG are highlighted in yellow and marked with *. We also noticed that the position of these features are correlated to the previously reported peak in the vicinity of the G band, assigned to superlattice-induced Raman process referred as R band [19]. For all measurements, we used at least two out of the three available laser excitation energies (E$_{laser}$) noticing that, different from the broadly studied DR Raman effect in graphene-based materials, their frequencies are E$_{laser}$ independent. In Fig. 2, the spectra are displayed according to the R frequency evolution; the correlation between the frequencies of the new families of Raman peaks highlighted in yellow and that of the R band is evident. All spectra were taken under resonance conditions by matching E$_{laser}$ with the energy interval between van Hove singularities in the electronic density of states, as reported in Refs. [22,23].

The origin of the new families of Raman peaks observed in our measurements is elucidated in the plots of Figure 3. In panel (a), the frequencies of all peaks in Figures 2 (a) and (b) highlighted in yellow and marked by * are plotted as a function of the R peak frequencies, which have been assigned as coming from the in-plane transversal optical (TO) phonon branch [19,20]. The inter-correlation between the * peaks frequencies is obvious, although different trends are observed for distinct families of peaks. Following the assumption that these new features are originated from other phonon branches, we have included the theoretical dispersion curves for the ZA, TA, LA, ZO, TO and LO branches [solid lines in Fig. 3(a)], as a function of the theoretical TO frequency

(according to the literature nomenclature: Z stands for out-of-plane modes; T and L stand for transversal and longitudinal at the zone center Γ, respectively; A and O stand for acoustic and optic, respectively). For that purpose, we have used the phonon dispersion curves calculated by Venezuela *et al.* [26], along the Γ-K direction (K stands for the Dirac point) in the first Brillouin zone. The actual phonon dispersion curves along the Γ-K direction are depicted in Fig. 3(b), and we have applied an up shift of 14 cm$^{-1}$ at the TO and LO frequencies to force the matching of the G band frequency at the Γ point ($\approx$1585 cm$^{-1}$). The assignment in Fig. 3(a) clearly shows that, while the R and R' bands (indicated in Fig. 2) come from the TO and LO phonon branches (respectively), as proposed before [19,20], most of the additional families of peaks come from other phonon branches predicted theoretically for monolayer graphene. However, the experimental data marked by open squares clearly deviates from the ZA graphene phonon branch and, as we will discuss later, they are related to the layer breathing vibrations, named ZO' phonons in AB stacked graphite.

As discussed by Carozo *et al.* [19], the R band frequency is related to a Raman process which involves the scattering of a photo-excited electron by a phonon with wavevector $\boldsymbol{q}$, and momentum conservation is achieved by the electron being elastically scattered by a superlattice wavevector -$\boldsymbol{q}$ determined by the rotational angle θ. Within the assumption that all other features present in the same spectrum are originated from similar processes, and thereby related to phonons with equal wavevector $\boldsymbol{q}$, we performed the assignment of the experimental data depicted in Fig. 3(a) in the phonon dispersion curves of monolayer graphene shown in Fig. 3(b). First, we assigned the R band frequency observed in the Raman spectrum of each specific sample to the theoretical TO phonon branch, thus obtained the respective wavevector $\boldsymbol{q}$. Then, we considered the same wavevector to assign all other features observed in the same spectrum. The result is depicted in Fig. 3(b), which shows measured frequencies (solid circles and open squares) as a function of $q$. The assignment of the experimental data to the phonon dispersion curves of monolayer graphene works well for most of the observed families of peaks, except for a low frequency family highlighted by different symbols (open squares). The origin of these low frequency peaks can be explained by considering layer-breathing mode vibrations (namely ZO' mode) occurring in bi-layer graphene, multi-layer graphene, and graphite. The overtone 2ZO' has been recently observed for few-layer graphenes (from two to six layers) and graphite in Reference [27]. The open black circles in Fig. 3(b) show experimental ZO' frequencies obtained for bi-layer graphene in Ref. [27], where distinct data points were taken using different laser energies, and the assignment between $E_{laser}$ and $q$ was based on the selection rule for achievement of usual DR condition [4,5]. Notice that these data points follow the same trend as our experimental data (open squares), and the dashed line in Fig. 3(b) is a tentative dispersion curve for the ZO' branch in tBLG graphene (along the Γ-K direction) that fits both datasets. The observation of the ZO' phonon mode in tBLG graphene reveals the interaction between the two rotationally-stacked planes.

According to the superlattice-induced Raman model applied to tBLG described in Ref. [19], the $q$ vector of the phonon involved in the scattering process is related to the rotational angle θ as

$$q = \frac{2\pi}{\sqrt{3}a}\{[-(1-cos\theta)-\sqrt{3}sin\theta]\hat{k}_x + [-\sqrt{3}(1-cos\theta)+sin\theta]\hat{k}_y\}, \quad (1)$$

where $a$ = 2.46 Å is the lattice parameter of graphene, and $\hat{k}_x$ and $\hat{k}_y$ are unit wavevectors, as defined in Figure 4.

Using equation (1), we were able to assign the frequencies of the observed Raman features to the rotational angle θ, and the results are depicted in Fig. 3(c). Discrete points are experimental Raman data, and solid lines are theoretical curves also obtained by assigning the theoretical curves shown in panel (b) to θ, using eq. (1). The ◊, x, ⊕, and △ symbols are experimental data obtained from References [19], [22], [23] and [28], respectively, where θ was independently determined by microscopy methods (high-resolution transmission electron microscopy [22,23,28], scanning tunneling microscopy [28] and lattice resolution atomic force microscopy [19]).

**Conclusions**

In summary, we have shown that the superlattice-induced Raman scattering allows us to probe the phonon dispersion of twisted bi-layer graphene (tBLG). The process is different from the broadly studied DR Raman process in graphene-based materials in many ways: (1) the phonon wavevectors $q$ in the interior of the Brillouin zone can be spanned without changing the excitation laser energy, but changing the twisting angle θ; (2) first-order scattering is activated by the superlattice modulation, without requiring defects to break the $q \cong 0$ selection rule; (3) the peak frequencies are defined by θ being $E_{laser}$ independent; (4) $q$ is not limited to wavevectors close to the high symmetry Γ and K points; (5) $q$ is a θ defined wavevector, including its direction, rather than an average of values around the high symmetry Γ and K points (see Fig. 4). The assignment provided here did not consider the details of the $q$-direction selective process. Deeper explorations can provide accurate experimental information for the development of theoretical models for the phonon dispersion in tBLG, including the coupling between layers in this system which has no AB or ABC stacking.

**Materials and Methods**

The studied tBLG graphene samples were produced by chemical vapor deposition (CVD) at low pressure using methane as carbon source and copper foil as catalyst [29]. 30 minutes annealing treatment is performed to the copper foil at 1000°C under an Ar/$H_2$ (10%) low flow. The synthesis is carried out at the same temperature by flowing $CH_4$ to the growth chamber during 40 minutes. After this time, the system is allowed to cool down, but the gases flux is maintained. In order to transfer the samples to Si/$SiO_2$ (100 nm-thick oxide) substrates, a standard transfer process is carried out using PMMA as support layer, and iron chloride ($FeCl_3$) to etch the copper. Raman spectroscopy measurements were performed on a LabRam Horiba instrument with excitation laser energies $E_{laser}$ = 2.54, 2.41 and 1.96 eV. A 1800 g/mm grating was used, and the laser power was maintained below 1 mW.

**Acknowledgements**

J.C.D. is thankful to CONACyT and FNRS for financial support. A.J and L.G.C. acknowledge financial support from CNPq and FAPEMIG.

FIGURES

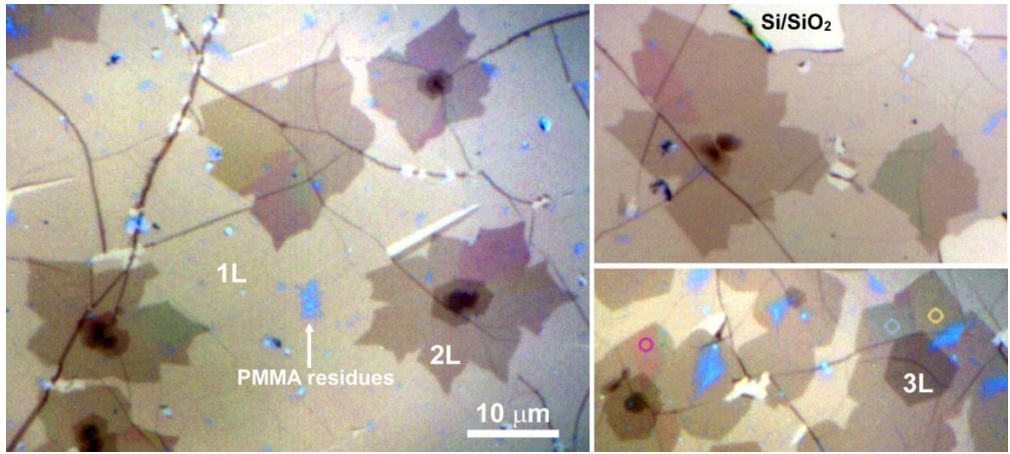

**Figure 1**. Representative optical micrographs of the studied sample transferred onto a Si/SiO$_2$ (100 nm) substrate (contrast enhanced), pink- blue- and yellow- colored areas can be evidenced in bi-layer graphene. Labels identify one-layer (1L), bi-layer (2L) and tri-layer (3L) areas as well as PMMA residues and Si/SiO$_2$. Pink, blue and yellow circles exemplify areas where the Raman measurements were recorded.

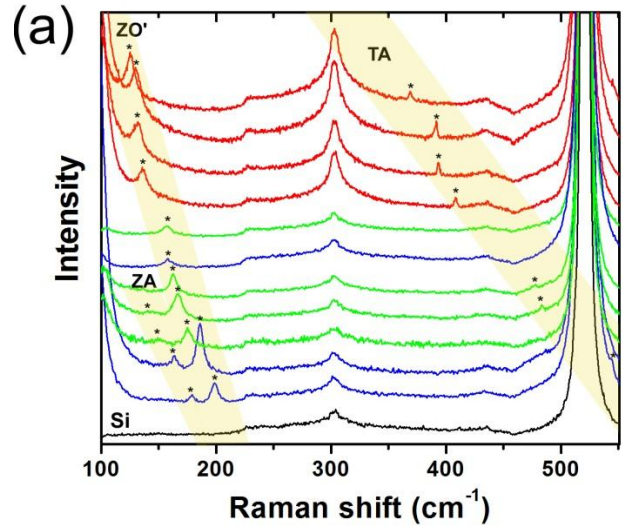

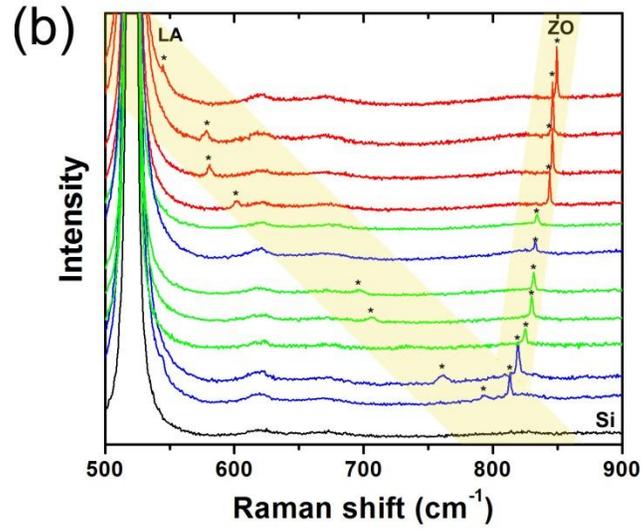

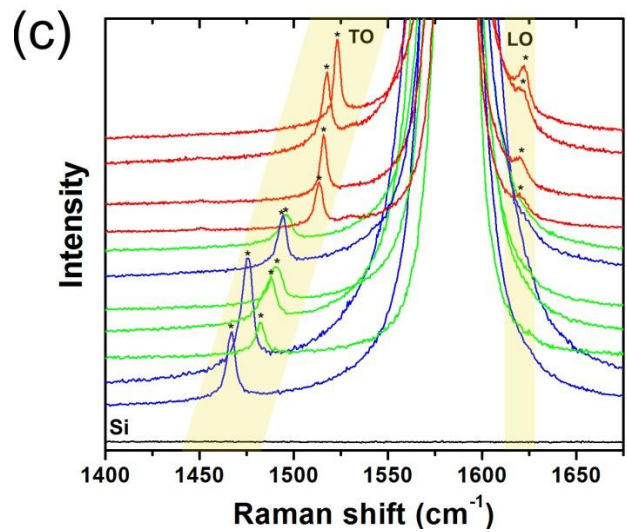

**Figure 2.** Raman spectra of CVD-grown tBLG with different rotational angles sitting on a Si/SiO$_2$ substrate. (a) 100-550 cm$^{-1}$ range, three families of new peaks can be observed and are highlighted in yellow; (b) 500-900 cm$^{-1}$ range, two new families of peaks (highlighted in yellow) can be observed; and (c) G band range where the R (TO) and R' (LO) bands are highlighted in yellow. Labels in the yellow areas correspond to the corresponding phonon branches assigned further in the text and * point to the Raman peaks considered in the study. Curve coloring corresponds to E$_{laser}$ used; red: E$_{laser}$=1.96 eV, green: E$_{laser}$=2.41 eV and blue: E$_{laser}$=2.54 eV; exception for the bottom (black) spectra (E$_{laser}$=2.41 eV), which correspond to Si substrate reference signal.

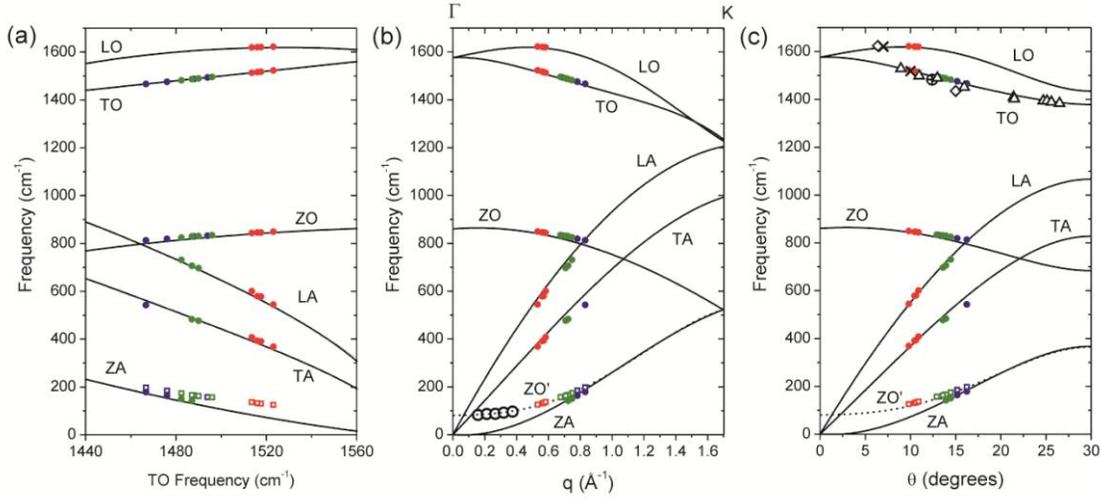

**Figure 3.** (a) Points indicate Raman frequencies of all the observed families of peaks from Fig. 2 as a function of R bands frequencies. Lines stand for the phonon frequencies (branch assignment near each curve) as a function of the TO phonon frequency along Γ-K, as obtained by Venezuela *et al.* [26]. Red, green and blue symbols stand for measurements obtained with E$_{laser}$ = 1.96, 2.41 and 2.54 eV, respectively. (b) Similar plot as in (a), but with frequencies plotted as a function of the wavevector *q*. The larger open black circles are data from Ref. [27]. (c) In this graph, the wavevectors *q* from the plot in (b) are replaced by the mismatch angles θ using equation (1). The ◇, x, ⊕, and △ symbols are experimental data obtained from References [19], [22], [23] and [28], respectively.

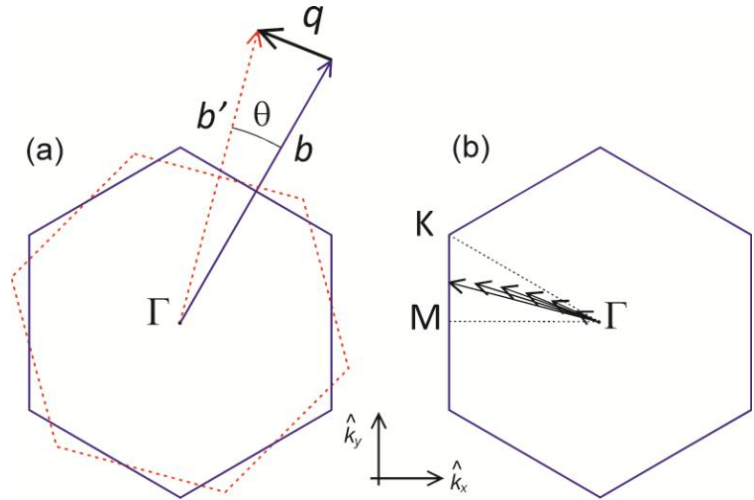

**Figure 4.** (a) Schematics defining the wavevector $q$ of the superlattice formed by a mismatch angle θ between the two layers. As indicated, $q=b'-b$, where $b$ and $b'$ are primitive reciprocal lattice vectors of the top (first Brillouin zone drawn by solid lines) and bottom (first Brillouin zone drawn by dashed lines) layers, respectively. (b) Schematics showing the evolution of some available $q$ wavevectors by varying θ between 0 and 30 degrees.